\documentclass[traditabstract]{aa} 
\usepackage{pstricks} 
\usepackage{graphicx}
\usepackage{natbib}
\usepackage[latin1]{inputenc} 
\usepackage{epsfig}
\usepackage{aas_macros}
\usepackage{amssymb,amsmath}

\psset{xunit=\columnwidth,yunit=0.85\columnwidth} 

\newcommand{\subscr}[1]{_\mathrm{#1}}

\newcommand{\doverdt}[1]{\frac{\partial #1}{\partial t}}

\begin{document}
\title{The Dynamical Origin of the Multi-Planetary System HD45364}
\author{Hanno Rein \inst{1} \and John C. B. Papaloizou \inst{1} \and Wilhelm Kley \inst{2}}
\titlerunning{The Dynamical Origin of HD45364}
\authorrunning{Rein, Papaloizou \and Kley}

\institute{	University of Cambridge, Department of Applied Mathematics and Theoretical Physics,
		Centre for Mathematical Sciences, Wilberforce Road, Cambridge CB3 0WA, UK, 
              \email{hr260@cam.ac.uk}\\
		\and
		University of T\"ubingen, Institute for Astronomy and Astrophysics,
		Auf der Morgenstelle 10, 72076 T\"ubingen, Germany
	       }

\abstract{
The recently discovered planetary system HD45364 which consists of a Jupiter and Saturn mass planet is
very likely in a 3:2 mean motion resonance. 
The standard scenario to form planetary commensurabilities is convergent migration of two planets embedded in a
protoplanetary disc. When the planets are initially separated by a period ratio larger than two,
convergent migration will most likely lead to a very stable 2:1 resonance for moderate migration rates.
To avoid this fate, formation of the planets close enough to prevent this resonance may be proposed.
However, such a simultaneous formation of the planets within a small annulus,
seems to be very unlikely.

Rapid type III migration of the outer planet crossing the 2:1 resonance is one possible way around this problem. 
In this paper, we investigate this idea in detail. 
We present an estimate for the required convergent migration rate and confirm this with N-body and hydrodynamical simulations. 
If the dynamical history of the planetary system had a phase of rapid inward migration that 
forms a resonant configuration, we predict that the orbital parameters of the two planets
are always very similar and hence should show evidence of that. 

We use the orbital parameters from our simulation to calculate a radial velocity curve and compare it to observations. 
Our model can explain the observational data as good as the previously reported fit. 
The eccentricities of both planets are considerably smaller and the libration pattern is different. 
Within a few years, it will be possible to observe the planet-planet interaction directly and thus distinguish
between these different dynamical states. 
}

\date{Submitted: 30 August 2009 - Revised: 26 October 2009 - Accepted: 26 October 2009 }
\maketitle

%%%%%%%%%%%%%%%%%%%%%%%%%%%%%%%%%%%%%%%%%%%%%%%%%%%%%%%%%%%%%%%%%%%%%%%%%%%%%%%%%
\section{Introduction}
Over 400 extrasolar planets have already been discovered \citep{exoplanet} and their diversity keeps 
challenging planet formation theory. For example, the recently discovered 
multi-planetary system HD45364 raises interesting questions about its formation history.

The planets have masses of $m_1=0.1906M_{\text{Jup}}$ and $m_2=0.6891M_{\text{Jup}}$ and are orbiting the star 
at a distance of $a_1=0.6813~\text{AU}$ and $a_2=0.8972~\text{AU}$, respectively \citep{CorreiaUdry2008}. 
The period ratio is close to $1.5$ and a stability analysis implies that the planets are deep 
inside a 3:2 mean motion resonance. The planets have most likely formed further out in cooler 
regions of the proto-stellar disc, as water ice which is an important ingredient for dust 
aggregation can only exist beyond the snow line which is generally assumed to be at radii 
larger than $2~\text{AU}$ \citep{SasselovLecar2000}.

It is then usually assumed that migration due to planet disc interactions has moved the planets 
closer to the star. Although the details of this process are still hotly debated, the existence 
of many resonant multiplanetary systems supports this idea. During migration the planets can get 
locked into a commensurabilities after which the planets migrate together with a constant period ratio. 
In such a resonance, one or more resonant angles are librating \citep[see e.g.][]{LeePeale01}. 

For the planetary system HD45364 this standard picture poses a new problem. Assuming that the 
planets have formed far apart from each other, the outcome for the observed masses after migration is almost always a 2:1 
mean motion resonance, not 3:2 as observed. The 2:1 resonance that forms is found to be extremely stable. 
One possible way around this is a very rapid convergent migration phase that passes quickly through the 2:1 resonance.  

In this paper we explore this idea quantitatively. The plan of the paper is as follows.
In section \ref{sec:formationnbody} we show, using N-body simulations, that the system 
always ends up in the 2:1 resonance assuming moderate migration rates. We present scenarios 
which result in formation of a 3:2 resonance after a rapid migration phase. 
In section \ref{sec:formationhydro} we perform hydrodynamic simulations with a variety of disk models in order to
explore the dependence on the physical setup.
In section \ref{sec:otherformationscenarios} we briefly discuss other formation scenarios.
We go on to compare the orbital parameters of our simulations with the observed radial velocity data in 
section \ref{sec:observation}. 

We find that the orbital parameters observed in our simulations differ from those estimated from a statistical fit
by \cite{CorreiaUdry2008}. However, our models reproduce the observational data very well and we argue
that our fit has the same level of significance as that obtained by \cite{CorreiaUdry2008}.

Future observations will be able to resolve this issue. 
This is the first prediction of orbital parameters for a specific extrasolar planetary system 
derived from planet migration theory only. 
The parameter space of orbital configurations produced by planet disc interactions (low eccentricities, 
relatively small libration amplitudes) is very small.
As in case of the GJ876 system, this can provide strong evidence on how the system formed.
Finally, we summarise our results in section \ref{sec:conclusion}.

%%%%%%%%%%%%%%%%%%%%%%%%%%%%%%%%%%%%%%%%%%%%%%%%%%%%%%%%%%%%%%%%%%%%%%%%%%%%%%%%%
\section{Formation of HD45364}
\label{sec:formationnbody}
\subsection{Convergent migration and resonance capture}
In the core accretion model \citep[for a review see e.g.][]{Lissauer1993} a solid core is firstly formed by dust aggregation. 
This process is much more efficient if water exists in solid form. In the proto-stellar nebula this happens
beyond the ice line where the temperature is below $150~\text{K}$ at distances larger than a few $\text{AU}$. 
Subsequently, after a critical mass is attained \citep{Mizuno1980}, the core accretes a gaseous envelope 
from the nebula \citep{BodenheimerPollack86}. Both planets in the HD45364 system are interior to the ice 
line, implying that they should have migrated inwards.

The migration rate depends on many parameters of the disc such as surface density and viscosity
as well as the mass of the planets. The planets are therefore in general expected to 
have different migration rates which leads to the possibility of convergent migration. 
In this process the planets approach orbital commensurabilities.
If they do this slowly enough, resonance capture may occur \citep{Gold65}
after which they migrate together maintaining a constant period ratio thereafter. 

\begin{figure}[tb]
\centering
\includegraphics[width=.95\columnwidth,trim=0.3cm 0 -0.2cm 0.5cm,clip=true]{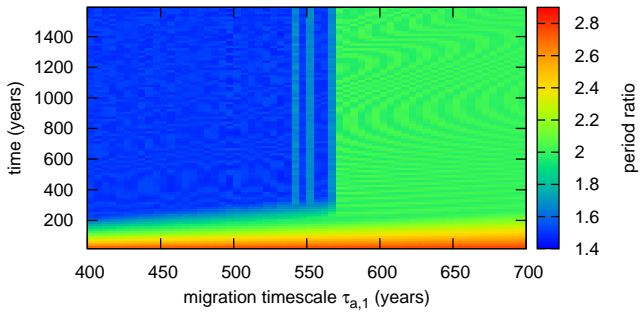}
\caption{Period ratio $P_2/P_1$ as a function of time ($y$-axis) and migration timescale 
of the outer planet $\tau_{a,2}$ ($y$-axis). The migration timescale of the inner planet 
is $\tau_{a,1}=2000~\text{yrs}$. 
The inner planet is initially placed at $r_1=1\text{AU}$.
The eccentricity damping is given through 
$K\equiv \tau_a / \tau_e =10$.	\label{fig:nbodym_taua2_2000_K_10}}
\end{figure}

Studies made by a number of authors have shown that when two planets, of either equal 
mass or with the outer one being the more massive, undergo differential convergent migration,
capture into a mean motion commensurability is expected to occur provided that
the convergent migration rate is not too fast \citep{SnellgrovePapaloizouNelson01}. 
The observed inner and outer planet masses are such that, if (as is commonly assumed for multiplanetary systems of this kind)
the planets are initially widely enough separated so that their period ratio exceeds $2,$
at low migration rates a 2:1 commensurability is expected to form
\cite[e.g.][]{NelsonPapaloizou2002,KleyPeitz04}.

\cite{PierensNelson2008} studied a similar scenario where the goal was to resemble 
the 3:2 resonance between Jupiter and Saturn in the early solar system. 
They also found that the 2:1 resonance forms at early stages.
However, in their case the inner planet had the larger mass whereas the planetary system
that we are considering has the heavier planet outside. 
In this situation the 2:1 resonance can be unstable, enabling the formation
of a 3:2 resonance later on and the migration rate may stall or even reverse \citep{MassetSnellgrove2001}.

%%%%%%%%%%%%%%%%%%%%%%%%%%%%%%%%%%%%%%%%%%%%%%%%%%%%%%%%%%%%%%%%%%%%%%%%%%%%%%%%%
\subsection{The 2:1 mean motion resonance}
\label{sec:formationnbody2:1}
We found that if two planets with masses of the observed system are in a 2:1 mean motion resonance, 
which has been form via convergent migration, this resonance is very stable.
An important constraint arises, because as indicated above, provided the planets start migrating
outside any low order commensurability, at the slowest migration rates a 2:1 resonance is expected to form
rather than the 3:2 commensurability that is actually observed.

We can estimate the critical relative migration timescale $\tau_{a,\text{crit}}$
above which a 2:1 commensurability forms from the condition that
the planets spend at least one libration period while migrating through the resonance.
The resonance semi-major axis width $\Delta a$ associated with the 2:1 resonance can be estimated from
the condition that two thirds of the mean motion difference across $\Delta a$
be equal in magnitude to $2\pi$ over the libration period. This gives
\begin{eqnarray}
\Delta a &=& \frac{\omega_{lf}a_2}{n_2} \label{eq:reswidth}
\end{eqnarray}
where $a_2$ and $n_2$ are the semi major axis and the mean motion of the outer planet, respectively. 
The libration period $2\pi/w_{lf}$ can be expressed in terms of the orbital parameters 
\citep[see e.g. ][]{Gold65, Rein2008} but is, for convenience, here measured numerically. 
If we assume the semi-major axes of the two planets
evolve on constant (but different) timescales $ |a_1/{\dot a_1}| = \tau_{a,1}$ 
and $|a_2/{\dot a_2}| = \tau_{a,2}$, the condition that the resonance width
is not crossed within a libration period gives
\begin{eqnarray}
\tau_{a,\text{crit}} \equiv \left| \frac{1}{1/\tau_{a,1}-1/\tau_{a,2}}\right| 
 & \gtrsim & 2\pi {{a_2} \over {\omega_{lf}\Delta a }} 
 = 2 \pi \frac{n_2}{\omega_{lf}^2 }
\end{eqnarray}
to pass through the 2:1 MMR. 

If the planets of the HD45364 system are placed in a 2:1 resonance with the inner planet 
being located at $1~\text{AU}$, the libration period $2\pi /\omega_{lf}$ is found to be 
approximately $75~\text{yrs}$. Thus, a relative migration timescale shorter than 
$\tau_{a,\text{crit}}\approx810~\text{yrs}$ is needed in order to pass through the 2:1 resonance. 
For example, if we assume that the inner planet migrates on a timescale of 2000 years, the outer 
planet has to migrate with a timescale
\begin{eqnarray}
\label{eq:analyticestimate} \tau_{a,2,crit} \lesssim 576~\text{yrs}. 
\end{eqnarray}

We have run several $N$-body simulations to explore the large parameter space and confirm the above estimate. 
The code used is similar to that presented in \cite{Rein2008} and uses a fifth order Runge-Kutta as well as 
a Burlish Stoer integrator, both with adaptive time-stepping. Different modules deal with migration and 
stochastic forcing. Non conservative forces are calculated according to the procedure presented in 
\cite{LeePeale2002} where the migration and eccentricity damping timescales $\tau_a= \left|{a}/{\dot a}\right|$ 
and $ \tau_e= \left| {e}/{\dot e}\right|$ are imposed for each planet individually. Stochastic forcing was employed 
only for one run in this paper and we refer the reader to \cite{Rein2008} for more details on the implementation. 

We place both planets on circular orbits at~$a_1~=~1~\text{AU}$ and $a_2~=~2~\text{AU}$ initially. 
The migration timescale for the inner planet is fixed at~$\tau_{a,1}=2000~\text{yrs}$, while the migration 
timescale for the outer planet is varied. In figure~\ref{fig:nbodym_taua2_2000_K_10} we plot the period 
ratio~$P_2/P_1$ as a function of time~$t$ for different migration timescales~$\tau_{a,2}$. There is a sharp 
transition of the final resonant configuration from 2:1 to 3:2 at around $\tau_{a,2}\approx 565~\text{yrs}$. 
Note that this value agrees extremely well with the analytic estimate.

The above results show that if the planets begin with a period ratio exceeding two, in order to get them 
into the observed 3:2 resonance, the relative migration time has to be shorter than what is obtained from 
the standard theory of type II migration applied to these planets in a standard model disc \citep{Nelson2000}.
In that case one expects this timescale to be $\sim 10^4 ~\text{yrs}$. However, it is possible to obtain the 
required shorter migration timescales in a massive disc in which the planets migrate in a type III regime 
\cite[see e.g.][]{MassetPapaloizou2003, Peplinski2008}. In this regime, the surface density distribution in the 
co-orbital region is asymmetric, leading to a large torque which is able to cause the planet to fall 
inwards on a timescale much shorter than the disc evolution time obtained for type II migration. 
We explore this possibility in more detail in the next section.

%%%%%%%%%%%%%%%%%%%%%%%%%%%%%%%%%%%%%%%%%%%%%%%%%%%%%%%%%%%%%%%%%%%%%%%%%%%%%%%%%%%%%%%%%%%%%5
\section{Hydrodynamical Simulations}
\label{sec:formationhydro}
 
\begin{table*}[tb]
\begin{center}
\begin{tabular}{l|lll|llll|llll|l}
\hline
\hline
   Run &  $10^4 q_1$& $10^4q_2 $& $r_2/r_1$  & $H/R$ & $b/H$& $\nu$ & $\Sigma$   & $r_i$ &$r_o$&  $N_r$ & $N_\phi$ & Result   \\ \hline
%\hline
\texttt{F1} & 2.18 & 7.89 &1.7 &0.05 &0.6 &$10^{-5}$ & 0.001   &0.25 &3.0& 768 & 768 &3:2\\%\hline	% conf_b0.60_h0.05_sigma0.0010_1cut
\texttt{F2} & 2.18 & 7.89 &1.7 &0.05 &0.6 &$10^{-5}$ & 0.0005  &0.25 &3.0& 768 & 768 &3:2 D\\%\hline	% conf_b0.60_h0.05_sigma0.0005_1cut
\texttt{F3} & 2.18 & 7.89 &1.7 &0.05 &0.6 &$10^{-5}$ & 0.00025 &0.25 &3.0& 768 & 768 &2:1 D\\%\hline	% conf_b0.60_h0.05_sigma0.00025_1cut
\texttt{F4} & 2.18 & 7.89 &1.7 &0.04 &0.6 &$10^{-5}$ & 0.001   &0.25 &3.0& 768 & 768 &3:2 D\\%\hline	% conf_b0.60_h0.04_sigma0.0010_1cut
\texttt{F5} & 2.18 & 7.89 &1.7 &0.07 &0.6 &$10^{-5}$ & 0.001   &0.25 &3.0& 768 & 768 &3:2\\%\hline	% conf_b0.60_h0.07_sigma0.0010_1cut
\texttt{R1} & 2.18 & 7.89 &1.7 &rad  &\textbf{1.0} &$10^{-5}$ & 0.0005  &0.25 &3.0& 300 & 300 &3:2\\\hline
\hline
\end{tabular}
\end{center}
\caption{Parameters for some of the hydrodynamic simulations:
The first column labels the run, the second and the third columns give the mass ratios
relative to the central mass of the inner and outer planets, respectively.
The fourth column gives the initial ratio of the semi-major axes of the two planets.
The fifth column gives the initial disc aspect ratio, the sixth the softening length in units of the
local scale height, the seventh the constant dimensionless viscosity and the eighth column gives the initial surface density.
The ninth and tenth columns give the inner and outer boundary of the 2D grid, respectively. 
The eleventh and twelfth columns give the radial and azimuthal grid resolutions used and finally 
the simulation outcome is indicated in the 
final column (see text). 
 \label{tab:sun1}}
\end{table*}

Two dimensional, grid based hydrodynamic simulations of two gravitationally interacting planets which 
also undergo interaction with an accretion disc have been performed in order to test the rapid migration hypothesis. 
The simulations performed here are similar in concept to those performed by \citet{SnellgrovePapaloizouNelson01} 
of the resonant coupling in the GJ876
system that may have been induced by orbital migration resulting from interaction with the protoplanetary disc.
We have performed studies using the FARGO code \citep{Masset00} with a modified locally isothermal equation of state. 
Those runs are indicated by the letter~\texttt{F}.

We have also run simulations including viscous heating and radiative transport using the {\tt RH2D} code, very similar to
\citet{KleyCrida08}. Those runs are indicated by the letter~\texttt{R}.

%%%%%%%%%%%%%%%%%%%%%%%%%%%%%%%%%%%%%%%%%%%%%%%%%%%%%%%%%%%%%%%%%%%%%%%%%%%%%%%%%%%%%%%%%%%%%5
\subsection{Equation of state}
The outer planet is likely to undergo rapid type III migration and the co-orbital region will be very asymmetric. 
We found, in accordance with \cite{Peplinski2008}, that the standard softening description does not lead to convergent results in 
this case \citep[for a comparative study see][]{Cridaetal2009}. 
Due to the massive disc, a large density spike develops near the planet. 
Any small asymmetry will then generate a large torque, leading to erratic results. 
We follow the prescription of \cite{Peplinski2008} and increase the sound speed near the outer planet since the locally 
isothermal model breaks down in the circumplanetary disc. The new sound speed is given by
\begin{eqnarray}
\label{eq:sound}
	c_s = \left[ \left( h_s r_s \right)^{-n} +\left( h_p r_p \right)^{-n} \right]^{-1/n} \cdot \sqrt{\,\Omega^2_s+\Omega^2_p\,}\label{eq:cs}
\end{eqnarray}
where $r_s$ and $r_p$ are the distance to the star and the outer planet, respectively. 
$\Omega_s$ and $\Omega_p$ are the Keplerian angular velocity and $h_s$ and $h_p$ are the aspect ratio, 
both of the circumstellar and circumplanetary disc, respectively. The parameter $n$ is chosen to be~$3.5$ 
and the aspect ratio of the circumplanetary disc is~$h_p=0.4$.
The sound speed has not been changed in the vicinity of the inner, less massive planet as it will not undergo 
type III migration and the density peak near the planet is much smaller. 

%%%%%%%%%%%%%%%%%%%%%%%%%%%%%%%%%%%%%%%%%%%%%%%%%%%%%%%%%%%%%%%%%%%%%%%%%%%%%%%%%%%%%%%%%%%%%5
\subsection{Radiation hydrodynamic simulations}
In the radiative simulations we go beyond the locally isothermal approximation
and include the full thermal energy equation, which takes into account the
generated viscous heat and radiative transport
\begin {equation}
\label{eq:energy}
 \doverdt{\Sigma c\subscr{v} T} + \nabla \cdot (\Sigma c\subscr{v} T {\bf u} )
       =  - p  \nabla \cdot {\bf u}  +  D  - Q  - 2 H \nabla \cdot \vec{F},
\end{equation}
where ${\bf u} = (u_r, u_{\varphi})$ is the two-dimensional velocity,
$\Sigma$ the surface density, $p$ the (vertically averaged) pressure, $T$ the (mid-plane)
temperature of the disc, and $c\subscr{v}$ is the specific heat at
constant volume. On the right hand side, the first term describes
compressional heating, $D$ the (vertically averaged) dissipation
function, $Q$ the local radiative cooling from the two surfaces
of the disc ($\propto T_{\rm eff}^4$), and $\vec{F}$ denotes the two-dimensional radiative flux
in the $(r, \varphi)$-plane. We calculate the effective temperature of the disc by 
$T_{\rm eff}^4 = \tau T^4$, where $\tau$ is the vertical optical depth of the disk.
To calculate $\tau$, we use a vertically averaged opacity.

For numerical stability we integrate the diffusion part implicitly. Further details
about solving the energy equation are stated in \citet{KleyCrida08}, and the code
{\tt RH2D} is described in more detail in \citet{1999MNRAS.303..696K}. 
In this radiative formulation the sound speed is a direct outcome of the simulations
and we do not use equation~\ref{eq:sound}.
 
%%%%%%%%%%%%%%%%%%%%%%%%%%%%%%%%%%%%%%%%%%%%%%%%%%%%%%%%%%%%%%%%%%%%%%%%%%%%%%%%%%%%%%%%%%%%%5
\subsection{Initial configuration and computational set up}

We use a system of units in which the unit of mass is the central mass $M_*,$ the unit of 
distance is the initial semi-major axis of the inner planet, $r_{1},$ and the unit of time 
is $(GM_*/r_{1}^3)^{-1/2},$ thus the orbital period of the initial orbit of the inner planet 
is $2\pi$ in these dimensionless units.

The parameters for some of the simulations we have conducted, as well as their outcomes
are given in table \ref{tab:sun1}, the caption being self-explanatory.

In all simulations a smoothing length $b=0.6~H$ has been adopted, where $H$ is the disc 
thickness at the planet position. This corresponds to about 4 zone widths in a simulation 
with resolution of $768\times768$. The role of the softening parameter $b$ that is used
in two dimensional calculations is to take account of the smoothing that would
result from the vertical structure of the disc in three dimensional calculations
(e.g. Masset et al 2006). We found that the migration rate of the outer planet is 
independent of the smoothing length if the sound speed is given by equation \ref{eq:cs}. 

The planets are initialised on circular orbits. The mass ratios adopted are
those estimated for HD45364. We assume there is no mass accretion so that these 
remain fixed in the simulation \cite[see discussion in][and also section \ref{sec:otherformationscenarios}]{Peplinski2008}.
The initial surface density profile is constant and tests indicate that
varying the initial surface density profile does not change the outcome very much.
Non-reflecting boundary conditions have been used throughout this paper.

%%%%%%%%%%%%%%%%%%%%%%%%%%%%%%%%%%%%%%%%%%%%%%%%%%%%%%%%%%%%%%%%%%%%%%%%%%%%%%%%%%%%%%%%%%%%%5
\subsection{Simulation results}

We found several possible simulation outcomes.
These are indicated in the final column of table \ref{tab:sun1} which 
gives the resonance obtained in each case with a final letter \texttt{D} denoting that
the migration became ultimately divergent resulting in the loss of commensurability. 

We found that convergent migration could lead to a 2:1 resonance that was set up in the initial 
stages of the simulation (\texttt{F3}).
Cases providing more rapid and consistent convergent 
migration than \texttt{F3} could attain a 3:2 resonance directly (\texttt{F1,F2,F4,F5}).

Thus avoidance of the attainment of sustained 2:1 commensurability
and the effective attainment of 3:2 commensurability required a rapid convergent
migration, as predicted by the N-body simulations and the analytic estimate 
(see section \ref{sec:formationnbody2:1}). That is apparently helped initially 
by a rapid inward migration phase of the outer planet that approaches a type III migration.
The outer planet went through that phase in all simulations with a surface density 
larger than $0.00025$ and therefore a 3:2 commensurability was obtained.
Thus, a surface density comparable to the minimum solar nebula \citep[MMSN,][]{Hayashi1981}, 
as used in simulation \texttt{F3}, is not sufficient to allow the planets undergo rapid enough type III migration.
As soon as the planets approach the 3:2 commensurability, type III migration 
stops due to the interaction with the inner planet and the outer planet starts to 
migrate in a standard type II regime. 

This imposes another constraint on the long term sustainability of the resonance. The inner planet 
remains embedded in the disc and thus potentially undergoes a fairly rapid inward type I migration. 
If the type II migration rate of the outer planet is slower than the type I migration rate of the inner planet, 
then the planets diverge and the resonance is not sustained. Note that a precise estimate of the 
migration rates is not possible at late stages, as the planets interact strongly with the 
density structure imposed on the disc by each other.

Accordingly, outflow boundary conditions at the inner boundary that prevent
the build up of an inner disc are more favourable to the maintenance of 
a 3:2 commensurability because the type I migration rate scales linearly with the disc 
surface density. However, those are not presented here, as the effect is small and 
we stop the simulation before a large inner disc can build up near the boundary.

The migration rate for the inner planet depends on the aspect ratio $h$. 
It is decreased for an increased disc thickness \citep{Tanakaetal2002}. 
That explains why models with a large disc thickness (\texttt{F5}) tend to stay in 
resonance whereas models with a smaller disc thickness (\texttt{F2,F4}) tend towards divergent migration at late times.
The larger thickness is consistent with radiative runs (see below).

\begin{figure}[htbp]
\centering
\includegraphics[width=0.8\columnwidth]{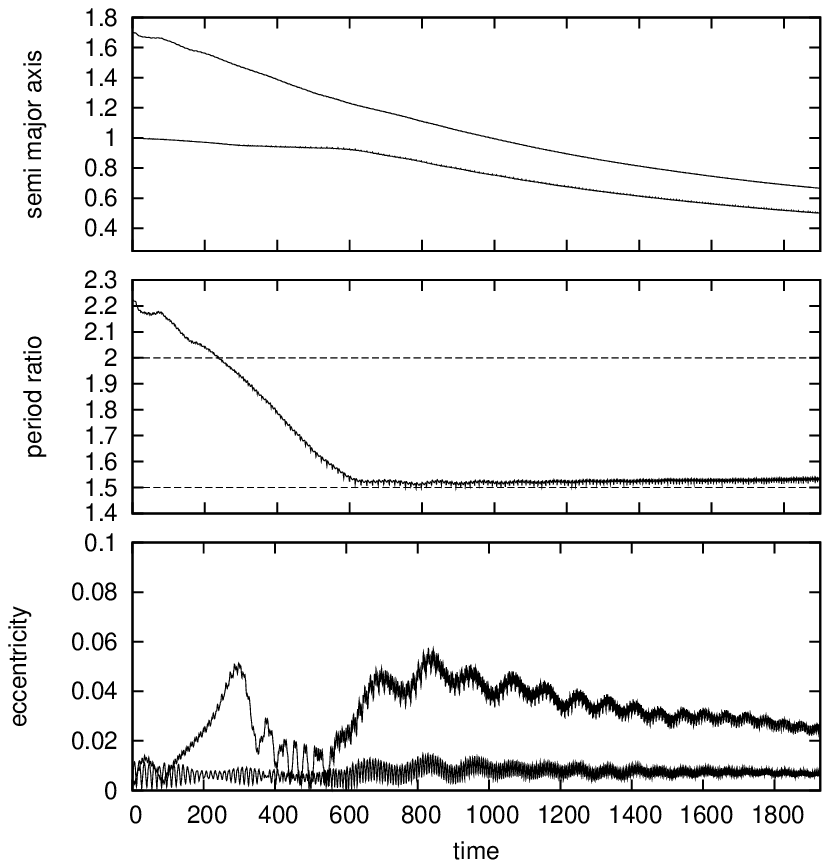}
\caption{The semi-major axes (top), period ratio $P_2/P_1$ (middle) and eccentricities (bottom) 
of the two planets plotted as a function of time 
in dimensionless units for run \texttt{F5} with disc aspect ratio of $h=0.07$. 
In the bottom panel the upper curve corresponds to the inner planet. 
\label{fig:conf_b0.60_h0.07_sigma0.0010_1cut_evolution}}
\centering
\begin{pspicture}(0,0)(1,0.90) 
\rput(0.5,0.47){\includegraphics[width=\columnwidth]{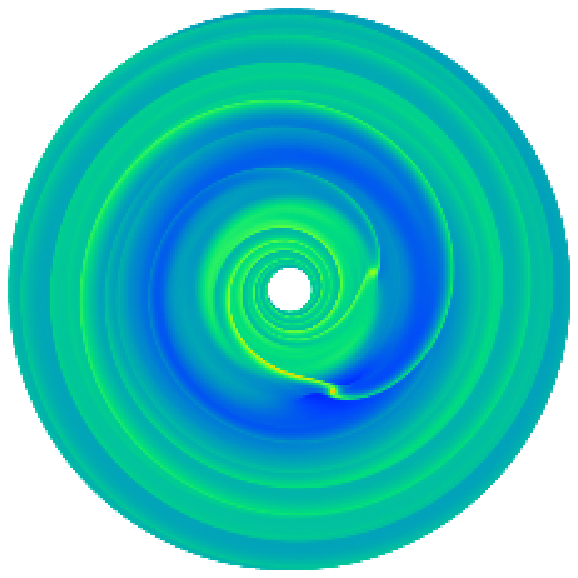}}
\rput(0.5,0.47){\includegraphics[width=\columnwidth]{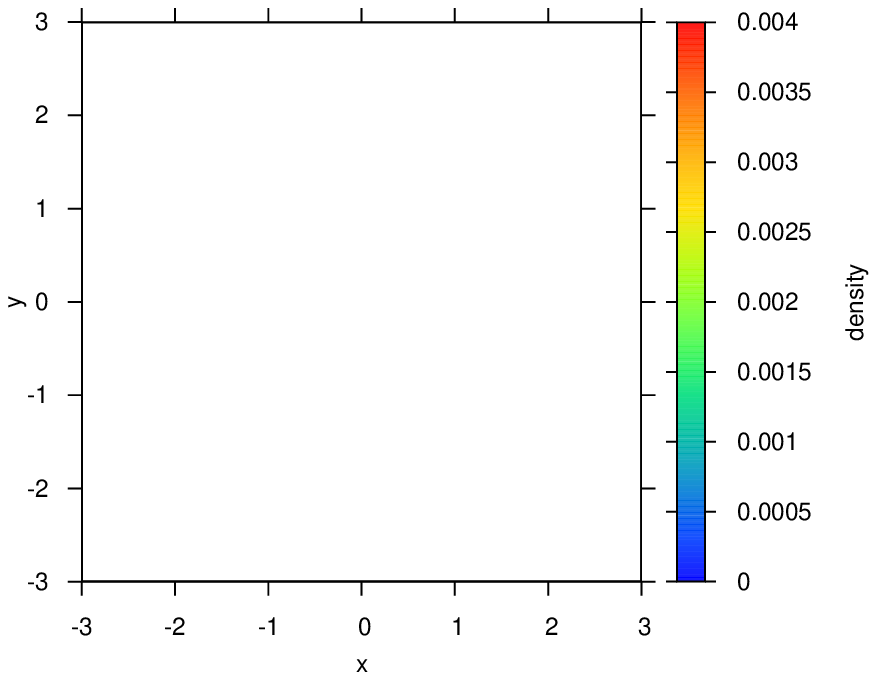}}
\end{pspicture}
\caption{A surface density contour plot for simulation \texttt{F5} after 
100 orbits at the end of the type III migration phase.
The outer planet establishes a definite gap while the inner planet
remains embedded at the edge of the outer planet's gap. 
\label{fig:conf_b0.60_h0.07_sigma0.0010_1cut_gasdens10}}
\end{figure}

As an illustration of the evolution from a typical configuration (\texttt{F5}) that forms and maintains
a 3:2 commensurability during which the orbital radii contract by a factor of at least
$\sim 2$ we plot the evolutions of the semi-major axes, the period ratio and eccentricities in figure
\ref{fig:conf_b0.60_h0.07_sigma0.0010_1cut_evolution} and provide a surface density
contour plot in figure \ref{fig:conf_b0.60_h0.07_sigma0.0010_1cut_gasdens10}
after $100$ initial inner planet orbits at which the 3:2 commensurability has just been established.

The eccentricity peak in figure \ref{fig:conf_b0.60_h0.07_sigma0.0010_1cut_evolution} at $t\sim300$ is due to
passing through the 2:1 commensurability. At $t\sim800$ the 3:2 commensurability is reached and maintained 
until the end of the simulation. 
The surface density in this simulation is approximately 5 times higher compared to the MMSN at $1~\text{AU}$ \citep{Hayashi1981}.

\begin{figure}[htbp]
\centering
\includegraphics[width=0.8\columnwidth]{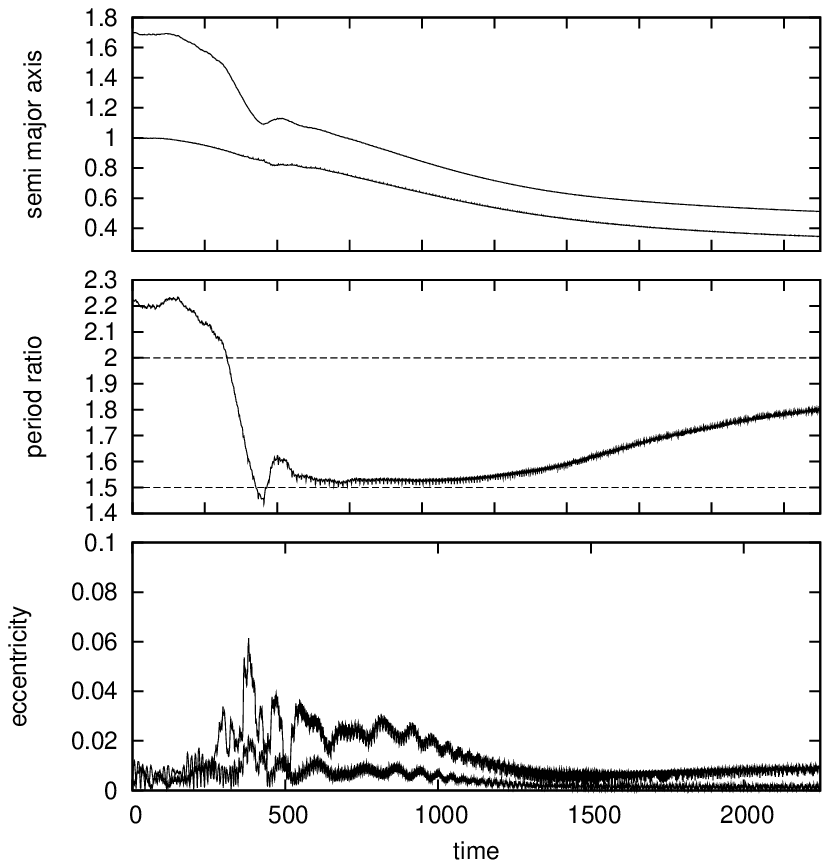}
\caption{The semi-major axes (top), period ratio $P_2/P_1$ (middle) and eccentricities (bottom) 
of the two planets plotted as a function of time 
in dimensionless units for run \texttt{F4} with a disc aspect ratio of $h=0.04$. 
In the bottom panel the upper curve corresponds to the inner planet. 
\label{fig:conf_b0.60_h0.04_sigma0.0010_1cut_evolution}}
\centering
\begin{pspicture}(0,0)(1,0.90) 
\rput(0.5,0.47){\includegraphics[width=\columnwidth]{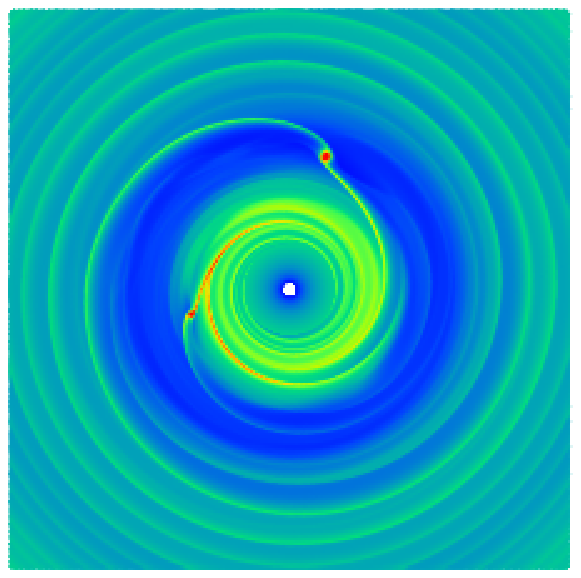}}
\rput(0.5,0.47){\includegraphics[width=\columnwidth]{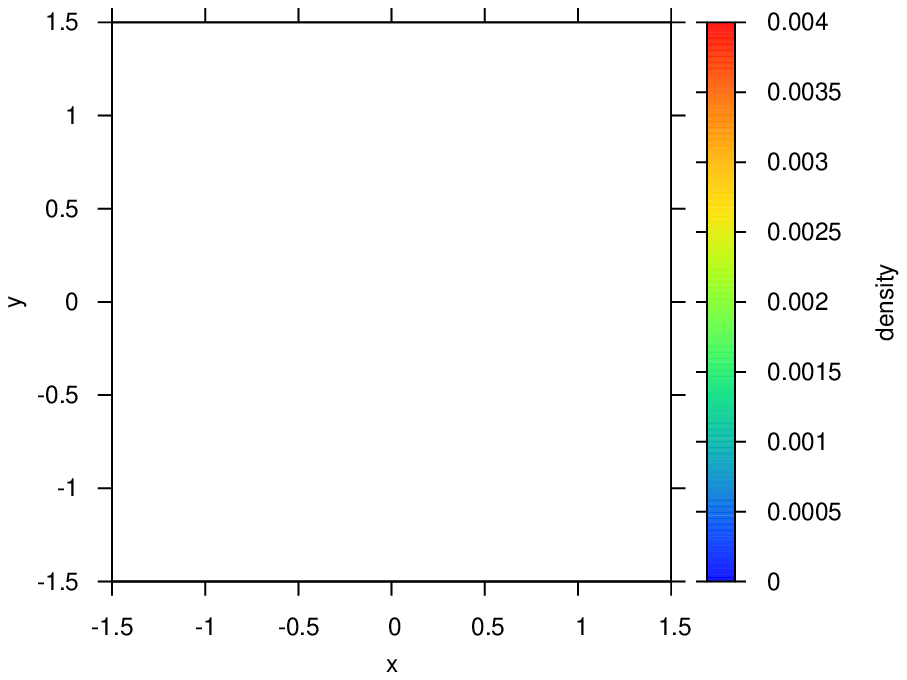}}
\end{pspicture}
\caption{A surface density contour plot for simulation \texttt{F4} after 
150 orbits after the planets went into divergent migration.
The inner planet is embedded and interacts strongly with the inner disc. 
The simulation uses a 1D grid for $0.04<r<0.25$.
\label{fig:conf_b0.60_h0.04_sigma0.0010_1cut_gasdens15}}
\end{figure}

We further present an illustration of the evolution of the semi major axes and eccentricities, 
as well as a surface density plot from a simulation (\texttt{F4}) that does form a 3:2 commensurability, 
but loses it because the inner planet is migrating too fast in figures
\ref{fig:conf_b0.60_h0.04_sigma0.0010_1cut_evolution} and 
\ref{fig:conf_b0.60_h0.04_sigma0.0010_1cut_gasdens15}.
One can see that a massive inner disc has been piled up. 
This, and and the small aspect ratio of $h=0.04$ makes 
the inner planet go faster than the outer planet which has opened a clear gap. 
The commensurability is lost at $t\sim1200$.

\begin{figure}[htbp]
\centering
\includegraphics[width=0.8\columnwidth]{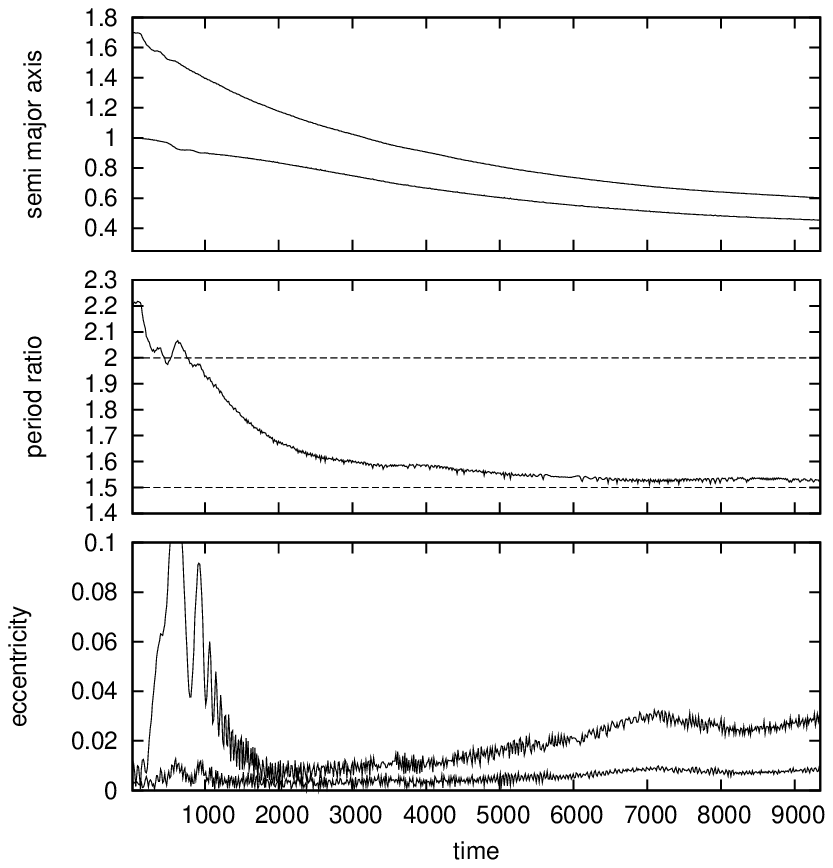}
\caption{The semi-major axes (top), period ratio $P_2/P_1$ (middle) and eccentricities (bottom) 
of the two planets plotted as a function of time in dimensionless units for the radiative run \texttt{R1}.
\label{fig:r1_evolution}}
\end{figure}

The radiative run \texttt{R1} also shows capture in a 3:2 resonance which is sustained.
Before embedding the planets the disc is first brought into radiative equilibrium
where viscous heating is balanced by radiative cooling. During this evolution
the surface density is kept constant as in the local isothermal models.
The temperature of the disc is obtained directly from the model, and the equilibrium
is equivalent to approximately $H/r=0.075$ at one AU. The planets are then embedded at the same
initial positions as in the other models (\texttt{F1-F5}). 
We plot the evolution of the semi major axes, period ratio and eccentricities in figure \ref{fig:r1_evolution}. 
When reaching the location of the 2:1 resonance the eccentricities of both planets increase
initially but decline subsequently after the resonance is transversed. Capture into 3:2 occurs
smoothly during the inward migration of the planets. The final eccentricities reached in this model
are around $e_1=0.03$ and $e_2=0.01$.

The initial type III migration rate is slower when compared to run \texttt{F5} because the
surface density is lower. Note that after the the 2:1 resonance is passed, the outer planet migrates 
in a type II regime and thus, the capture into the 3:2 resonance appears later. However, the orbital parameters
measured at the end of the simulations are very similar to any other run that we have performed. 
This indicates that the parameter space that is populated by these kind of planet disc simulation is very generic.

We tested the effect of 
disc dispersal at the late stages in our models to evolve the system self-consistently to the present day. 
In model \texttt{F5} after $t=2000$ we allow the disc mass to exponentially decay on 
a timescale of $\tau_{dis}\sim2000$. Note that this timescale is shorter than the 
photo-evaporation timescale \citep{AlexanderClarke2006}. However, this scenario is expected to give a stronger 
effect than a long timescale \citep{SandorKley06}. In agreement with those authors, we
found that the dynamical state of the system does not change for the above parameters. At a late stage, the resonance
is well established and the planets undergo a slow inward migration. Strong effects are only expected if the disc dispersal 
happens during the short period of rapid type III migration, which is very unlikely. We observed that the 
eccentricities show a trend to decrease and the libration amplitudes to slightly increase during the dispersal phase.  
However, these changes are not different than what has been observed in runs without a disappearing disc.

%%%%%%%%%%%%%%%%%%%%%%%%%%%%%%%%%%%%%%%%%%%%%%%%%%%%%%%%%%%%%%%%%%%%%%%%%%%%%%%%%%%%%%%%%%%%%5
\section{Other scenarios for the origin of HD45364}
\label{sec:otherformationscenarios}

In the above discussion we have considered the situation when the planets
attain their final masses while having a wider separation than that required 
for a 2:1 commensurability and found that convergent migration scenarios can 
be found that bring them into the observed 3:2 commensurability by disc planet interaction.
However, it is possible that they could be brought to their
current configuration in a number of different ways as considered below.
Nonetheless it is important to note that because the final commensurable
state results from disc planet interactions
it should have similar properties as those described above when making
comparisons with observation.

It is possible that the solid cores of, either both planets,
or just the outer planet approached the inner planet more closely than 
the 2:1 commensurability before entering the rapid gas accretion phase 
and attaining their final masses prior to entering the 3:2 commensurability.
Although it is difficult to rule out such possibilities entirely, we note 
that the cores would be expected to be in the super earth 
mass range where in general closer commensurabilities than 
2:1 and even 3:2 are found for typical type I migration rates
\citep[e.g.][]{PapaloizouSzuszkiewicz2005, CresswellNelson2008}.
One may also envisage the possibility that the solid cores grew in situ 
in a 3:2 commensurability, but this would have to survive expected strongly 
varying migration rates as a result of disc planet interactions as the 
planets grew in mass.

An issue is whether the embedded inner planet is in a rapid accretion phase.
The onset of the rapid accretion phase (also called phase  3) occurs when the core
and envelope mass are about equal \citep{Pollacketal1996}. The total planet
mass depends at this stage on the boundary conditions, here determined by the circumplanetary flow.
When these allow the planet to have a significant convective envelope the transition
to rapid accretion may not occur until the planet mass exceeds $60 \text{M}_{\oplus}$
\citep{Wuchterl1993}, which is the mass of the inner planet (see also model J3 of 
\citeauthor{Pollacketal1996} \citeyear{Pollacketal1996}, 
and models of \citeauthor{PapaloizouTerquem1999} \citeyear{PapaloizouTerquem1999}). 
Because of the above results, it is not unreasonable 
that the inner planet is not in a rapid accretion phase.

Finally we remark that protoplanetary discs are believed to maintain turbulence in some part
of their structure. Using the prescription of \cite{Rein2008}, we can simulate the turbulent 
behaviour of the disc by adding additional forces to an N-body simulation. 
These forces will ultimately eject the planets from the 2:1 resonance should that form. 
Provided they are strong enough this can happen within the lifetime of the disc, 
thus making a subsequent capture into the observed 3:2 resonance possible. 

We have confirmed numerically that such cases can occur for moderately large diffusion coefficients 
\citep[as estimated by][]{Rein2008}. However, this outcome seems to be the exception rather than the rule. 
Should the 2:1 resonance be broken, a planet-planet scattering event appears to be more likely. 
In all the simulations we performed, we found that only a small 
fraction of systems ($1\%$~-~$5\%$) eventually end up in a 3:2 resonance.

%%%%%%%%%%%%%%%%%%%%%%%%%%%%%%%%%%%%%%%%%%%%%%%%%%%%%%%%%%%%%%%%%%%%%%%%%%%%%%%%%
\section{Comparison with observations} \label{sec:observation}

\begin{table*}[htbp]
\begin{center}
\begin{tabular}{ll|ll|ll}
\hline\hline
& &\multicolumn{2}{l|}{\cite{CorreiaUdry2008}} & \multicolumn{2}{l}{Simulation \texttt{F5}}  \\
Parameter& Unit & b & c  & b & c \\\hline
$M \sin i$ & $[\mbox{M}_{\text{Jup}}]$ 		& 0.1872 & 0.6579 			& 0.1872 & 0.6579 		 		\\
$M_*$ &  $[M_{\odot}]$ 				&\multicolumn{2}{c|}{0.82} 		& \multicolumn{2}{c}{0.82} 			\\
$a$ &$[\mbox{AU}]$				& 0.6813 & 0.8972 			& 0.6804 & 0.8994  	 	\\
$e$ 	&		& 			$ 0.17\pm 0.02$ & $0.097 \pm0.012$ 	& 0.036  & 0.017  		\\
$\lambda$ & $[\mbox{deg}]$ 			& $105.8 \pm1.4$ & $269.5\pm0.6$ 	& 352.5 & 153.9 		\\
$\varpi$ & $[\mbox{deg}]$ 			& $162.6 \pm6.3$ & $7.4\pm4.3$ 		& 87.9& 292.2		\\
$\sqrt{\chi^2}$ &				& \multicolumn{2}{c|}{2.79} 		& \multicolumn{2}{c}{2.76$^\star$ (3.51)} 	\\
Date  &[JD]					& \multicolumn{2}{c|}{2453500} 		& \multicolumn{2}{c}{2453500}  	\\
\hline
\hline
\end{tabular}
\end{center}
\caption{Orbital parameters of HD45364b and HD45364c. 
The observed values including error-bars are taken from \cite{CorreiaUdry2008}. 
Note that $\varpi$ is not well constrained for nearly circular orbits.
The radial velocity amplitude of the simulation marked with $^\star$ has been scaled down by $8\%$.
\label{tab:orbit}}
\end{table*}

\begin{figure*}[tbp]
\centering
\includegraphics[width=1.0\textwidth]{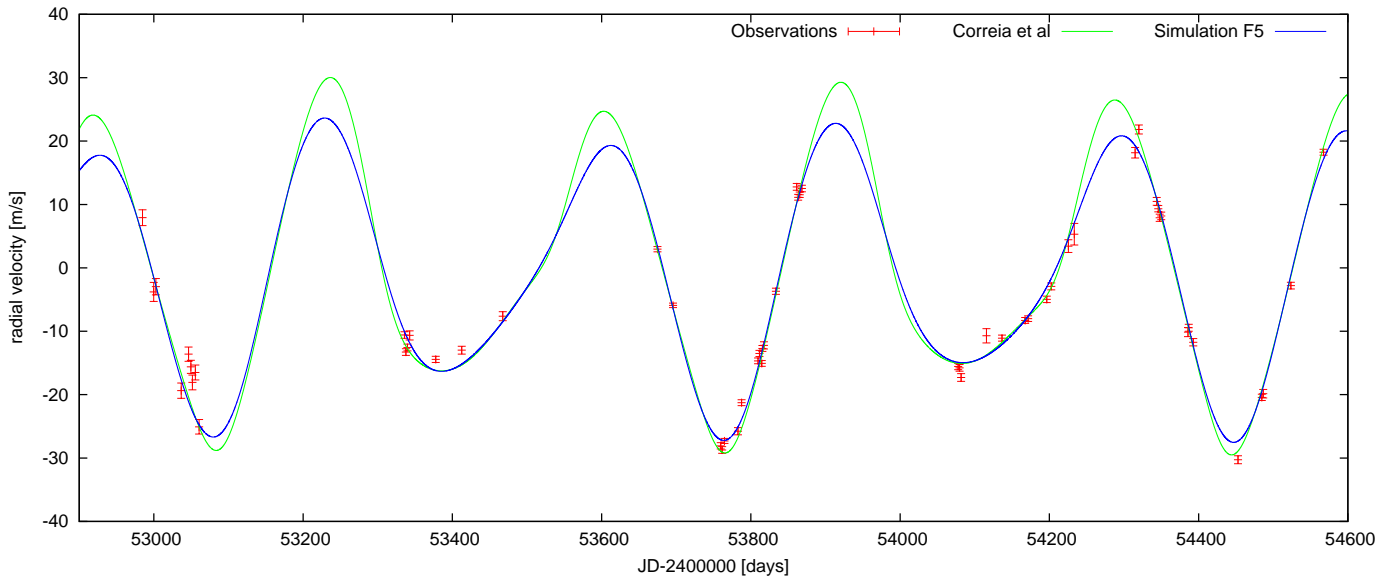}
\caption{Comparison of different orbital solutions. 
See text for a description of the different models. 
Radial velocity measurements (red points) and the published orbital 
solution (green curve) are taken from \cite{CorreiaUdry2008}.
\label{fig:nbodym_rv1}}
\end{figure*}

Table \ref{tab:orbit} lists the orbital parameters that \cite{CorreiaUdry2008} obtained 
from their best statistical fit of two Keplerian orbits to the radial velocity data. 
The radial velocity data collected so far is insufficient to detect
the presence of any interactions between the planets. 
This is borne out by the fact that the authors did not obtain an improvement in terms of
minimising $\chi^2$ when a 3-body Newtonian 
fit rather than a Keplerian fit was carried out. 
However, a stability analysis supports the viability of the determined
parameters as these lie inside a stable region of orbital parameter space.
The best fit solution shows a 3:2 mean motion commensurability
although no planet-planet interactions have been observed from which 
this could be inferred directly. 
It is important to note that the large minimum $\chi^2$ associated with 
the best fit that could be obtained indicates large uncertainties 
which are not accounted for by the magnitudes of the errors quoted in their paper. 

There are two main reasons why we believe that the effective
errors associated with the observations are larger:
First, many data points are clustered and clearly do not 
provide a random time sampling. This could result in 
correlations which effectively reduce the total number of independent measurements.
Second, the large minimum $\chi^2$ value being three times the quoted
observational error indicates that there are additional effects
(e.g. sunspots, additional planets, etc) that produce a jitter in the central star
(M. Mayor, private communication) which then enhances the effective observational error. 

No matter what process generates the additional noise, 
we can assume in a first approximation that 
it follows a Gaussian distribution. We then have to 
conclude that the effective error of each observation is a factor 
$\sim3$ larger than reported in order to account for the 
minimum $\chi^2$ given the quoted number
of $\sim 60$ independent observations. Under these circumstances
we would then conclude that any fit with $\sqrt{\chi^2}$
in the range $2-4$ would be an equally valid possibility.
 
In the following, we show that indeed a variety of orbital 
solutions match the observed RV data with no statistically significant difference
when compared to the quoted best fit by \cite{CorreiaUdry2008}. 
As one illustration, we use the simulation \texttt{F5} obtained in section \ref{sec:formationhydro}. 
We take the orbital parameters at a time when the orbital period 
of the outer planet is closest to the observed value 
and integrate them for several orbits with our N-body code. 
The solution is stable for at least one million years. 
There are only two free parameters available
to fit the reflex motion of the central star to the observed radial velocity: 
the origin of time (epoch) and the angle between the line of sight and the pericentre of the planets. 

We can assume that the planet masses and periods are measured with high accuracy and that 
only the shape of the orbit contains large errors. Our best fit results in an unreduced 
$\sqrt{\chi^2}$ value of $3.51$. According to the above discussion, this solution is statistically 
indistinguishable from a solution with $\sqrt{\chi^2}\sim2.8$.

It is possible to reduce the value of $\chi^2$ even further, when assuming that the planet masses
are not fixed. In that case, we have found a fit corresponding to $\sqrt{\chi^2}=2.76$ where we have reduced the
radial velocity amplitude by $8\%$ (effectively adding one free parameter to the fit). 
This could be explained either by a heavier star, less massive planets or a less inclined orbit.
The orientation of the orbital plane has been kept fixed ($i=90^\circ$) in all fits.

The results are shown in figure \ref{fig:nbodym_rv1}. 
The blue curve corresponds to the outcome of simulation \texttt{F5} and 
we list the orbital parameters in table \ref{tab:orbit}. 
We also plot the best fit given by \cite{CorreiaUdry2008} for comparison (green curve). 
In accordance with the discussion given above it is very 
difficult to see any differences in the quality of these fits 
which indeed suggests that the models are statistically indistinguishable. 

However, there is an important difference between
the fits obtained from our simulations and that of \cite{CorreiaUdry2008}.
Our models consistently predict lower values for both 
eccentricities $e_1$ and $e_2$ (see table \ref{tab:orbit}). 
Furthermore the ratio of eccentricities $e_1/e_2$ is higher than the previously reported value of $1.73$. 
Note however, that the eccentricities are oscillating and the ratio is on average $\sim3$.
 
This in turn results in a different libration pattern: the slow libration mode 
\citep[see figure 3a in][]{CorreiaUdry2008} that is associated
with oscillations of the angle between the two apsidal lines is absent, as shown 
in figure \ref{fig:nbody_angles}. 
Thus there is a marked difference in the form the interaction between the
two planets takes in the case of this fit.

\begin{figure}[tbp]
\centering
\includegraphics[angle=270,width=0.97\columnwidth]{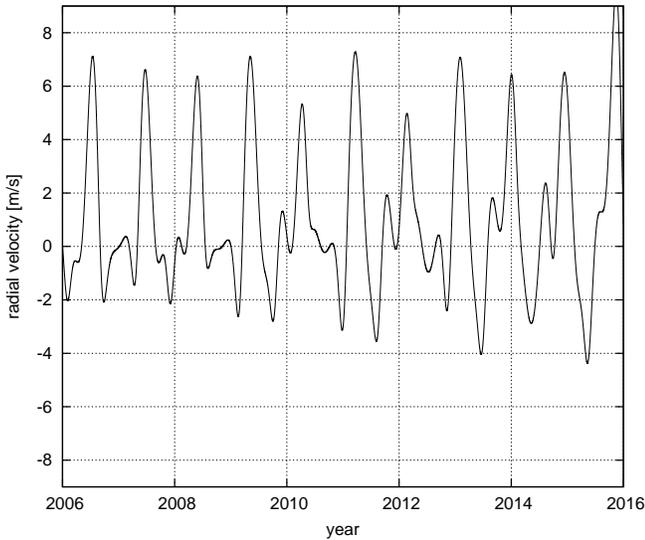}
\caption{Difference in radial velocity between the previously reported fit by 
\cite{CorreiaUdry2008} and the new fit obtained from simulation \texttt{F5}.
\label{fig:longtermdiff}}
\end{figure}

\begin{figure}[tbp]
\centering
\includegraphics[angle=270,width=0.99\columnwidth]{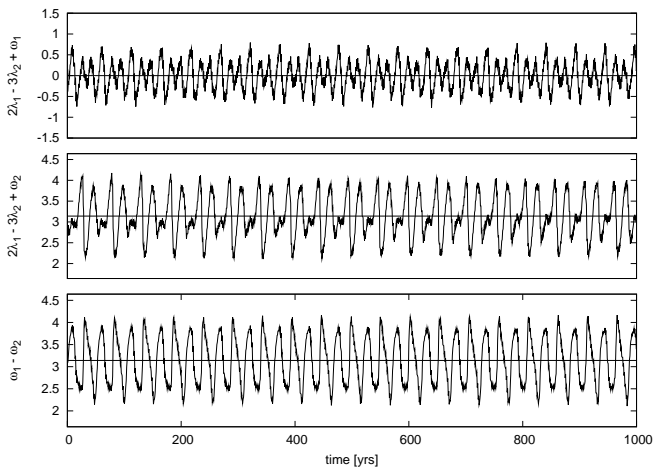}
\caption{Evolution of the resonant angles 
$\phi_1=2\lambda_1-3\lambda_2+\varpi_1$, $\phi_2=2\lambda_1-3\lambda_2+\varpi_2$, $\varpi_1-\varpi_2$ 
for simulation \texttt{F5} using the parameters given in table \ref{tab:orbit}. The angles are measured in radians.
\label{fig:nbody_angles}}
\end{figure}

It is hoped that future observations will be able to resolve this issue. 
The evolution of the difference in radial velocity over the next few years between our fit from simulation \texttt{F5} 
and the fit found by \cite{CorreiaUdry2008} is plotted in figure \ref{fig:longtermdiff}.
There is approximately one window each year that will allow further 
observations to distinguish between the two models. The large difference 
at these dates is not due to the secular evolution of the system, but simply a
consequence of smaller eccentricities.  


All runs that yield a final configuration with a 3:2 MMR are very
similar in their dynamical behaviour, as found for model \texttt{F5}. They are characterised
by an antisymmetric ($\varpi_2 - \varpi_1 \sim \pi$) state with relatively small
libration of $\phi_1$ and $\phi_2$. This is a generic outcome of
convergent disk-planet migration scenarios.

%%%%%%%%%%%%%%%%%%%%%%%%%%%%%%%%%%%%%%%%%%%%%%%%%%%%%%%%%%%%%%%%%%%%%%%%%%%%%%%%%
\section{Conclusions}
The planets in the multi-planetary system HD45364 are most likely in a 3:2 mean motion resonance. 
This poses interesting questions on it's formation history. 
Assuming that the planets form far apart from each other and migrate with 
moderate migration rate, as predicted by standard planet formation and 
migration theories, the most likely outcome is a 2:1 mean motion resonance, 
contrary to the observation of a 3:2 MMR.

In this work, we investigated a possible way around this problem by letting 
the outer planet undergo a rapid inward type~III migration.
We presented an analytical estimate and performed N-body as well as hydrodynamical simulations.

We found that it is indeed possible to form a 3:2 MMR and avoid the 2:1 resonance, 
thus resembling the observed planetary system using reasonable disc parameters. 
Hydrodynamical simulations suggest that the system is more likely to sustain 
the resonance for large aspect ratios, as the migration of the inner planet is slowed down, 
thus avoiding divergent migration. 

Finally, we used the orbital configuration found in the hydrodynamical 
formation scenario to calculate a radial velocity curve. 
This curve has then been compared to observations and the resulting fit 
has an identical $\chi^2$ value than the previously reported \textit{best fit}.

Our solution is stable for at least a million years. It is in a dynamically 
different state, both planets having lower eccentricities and a different libration pattern.
This is the first time that planet migration theory can predict a precise orbital 
configuration of a multiplanetary system. 

This might also be the first direct evidence for type III migration if 
this scenario turns out to be true. 
The system HD45364 remains an interesting object for observers, as the 
differences between the two solutions can be measured in radial velocity within a couple of years.

\label{sec:conclusion}

\begin{acknowledgements}
The authors are grateful to the Isaac Newton Institute for Mathematical Sciences in 
Cambridge where the final stages of this work were carried out during the Dynamics of Discs and Planets research programme.
Hanno Rein was supported by an Isaac Newton Studentship, the Science and 
Technology Facilities Council and St John's College Cambridge. 
Simulations were performed on the Astrophysical Fluid's Core i7 machines at DAMTP and on Darwin, 
the Cambridge University HPC facility.
\end{acknowledgements}

\bibliography{full}
\bibliographystyle{aa}

\end{document}